\documentclass{Rinton-P9x6}

\begin{document}

\def\cstok#1{\leavevmode\thinspace\hbox{\vrule\vtop{\vbox{\hrule\kern1pt
\hbox{\vphantom{\tt/}\thinspace{\tt#1}\thinspace}}
\kern1pt\hrule}\vrule}\thinspace}

\title{Photon Green Functions in Curved Spacetime}

\author{Giuseppe Bimonte, Enrico Calloni, 
Luciano Di Fiore,
{\underline {Giampiero Esposito}}, Leopoldo Milano, 
Luigi Rosa}

\address{Dipartimento di Scienze Fisiche, Universit\`a Federico
II di Napoli, Complesso Universitario di Monte S. Angelo,
Via Cintia, Edificio N', 80126 Napoli, Italy\\
INFN, Sezione di Napoli,
Complesso Universitario di Monte S. Angelo,
Via Cintia, Edificio N', 80126 Napoli, Italy}

\maketitle

\abstracts{Quantization
of electrodynamics in curved space-time in the Lorenz
gauge and with arbitrary gauge parameter makes it necessary to
study Green functions of non-minimal operators with variable
coefficients. Starting from the integral representation of photon
Green functions, we link them to the evaluation of integrals
involving $\Gamma$-functions. Eventually, the full asymptotic
expansion of the Feynman photon Green function at small values
of the world function, as well as its explicit dependence on
the gauge parameter, are obtained without adding by hand a mass
term to the Faddeev--Popov Lagrangian. Coincidence limits of second
covariant derivatives of the associated Hadamard function are also 
evaluated, as a first step towards the energy-momentum tensor
in the non-minimal case.}

\section{Introduction}

Recent investigations of the force acting on a rigid Casimir apparatus
in a gravitational field~\cite{ca} \cite{cd}  
have provided another relevant
example of how Green function methods~\cite{do}  
can prove useful in modern
gravitational physics. This has motivated our interest in a thorough
analysis of the photon Green functions in curved space-time, starting
with the case of curved manifolds without boundary. Our results are
a concise formula for the local asymptotics of the Feynman photon
Green function in the case of a non-minimal operator acting on the
electromagnetic potential, and an expression for second covariant
derivatives of the associated Hadamard function in the coincidence
limit. The latter can be used, in turn, to study the regularized
energy-momentum tensor. All of our analysis is performed 
by using $\zeta$-function regularizaton without
adding by hand a mass term to the BRST-invariant action, which is
indeed desirable from the point of view of general principles in
quantum field theory: a mass term added by hand spoils gauge invariance
of the Maxwell part of the action, nor does it respect BRST invariance
of the full action. Moreover, the addition by hand of a term 
${m^{2}\over 2}A_{\mu}A^{\mu}$ leads to a bad ultraviolet behaviour
of the photon propagator in momentum space. This is taken care of
by introducing an auxiliary vector field, but then unitarity of the
quantum theory is lost.~\cite{es}

\section{From the heat kernel to the Green function}

We consider quantum Maxwell theory in curved space-time via path
integrals, hence adding gauge-fixing term and ghost-fields
(here $\chi,\psi$) contribution. The full action reads therefore
as (hereafter $g=-{\rm det} \; g_{\mu \nu}(x)$)
\begin{equation}
S=\int d^{4}x \sqrt{g}\left[-{1\over 4}g^{\mu\rho}g^{\nu \beta}
F_{\mu \nu}F_{\rho \beta}
-{(\nabla^{\mu}A_{\mu})^{2}\over 2\alpha}
-{\chi \over \sqrt{\alpha}} \cstok{\ } \psi \right].
\label{(1)}
\end{equation}
This leads to the gauge-field operator
\begin{equation}
P^{\mu \nu}(\alpha)=-g^{\mu \nu}\cstok{\ } +R^{\mu \nu}
+\left(1-{1\over \alpha}\right)\nabla^{\mu}\nabla^{\nu},
\label{(2)}
\end{equation}
with associated 'heat kernel' obeying the first-order equation
\begin{equation}
{\rm i}{\partial \over \partial \tau}K_{\mu \nu'}^{(\alpha)}(\tau)
=P_{\mu}^{\; \lambda}(\alpha)K_{\lambda \nu'}^{(\alpha)}(\tau),
\label{(3)}
\end{equation}
with initial condition $K_{\mu \nu'}^{(\alpha)}(\tau=0)=
g_{\mu \nu}(x)\delta(x,x')$. Thanks to the work 
of Endo,~\cite{en} the
heat kernel for arbitrary values of $\alpha$ can be obtained from
the heat kernel at $\alpha=1$ through the formula
\begin{equation}
K_{\mu \nu'}^{(\alpha)}(\tau)=K_{\mu \nu'}^{(1)}(\tau)
+{\rm i} \int_{\tau}^{\tau / \alpha}dy \nabla_{\mu}
\nabla^{\lambda}K_{\lambda \nu'}^{(1)}(y).
\label{(4)}
\end{equation}
The Feynman photon Green function is eventually obtained from the
definition
\begin{equation}
g^{{1\over 4}}G_{\mu \nu'}^{(\alpha)}{g'}^{{1\over 4}}
\equiv \lim_{s \to 0}{\mu_{A}^{2s}{\rm i}^{s+1}\over \Gamma(s+1)}
\int_{0}^{\infty}d\tau \; \tau^{s}K_{\mu \nu'}^{(\alpha)}(\tau),
\label{(5)}
\end{equation}
where the limit as $s \rightarrow 0$ should be taken {\it at the
end of all calculations}, and the regularization involving 
$\tau^{s}$ is necessary since we do not rotate the integration
contour nor do we add by hand the ${m^{2}\over 2}A_{\mu}A^{\mu}$ 
term to the action as we said before. By virtue of Eqs. (4) and
(5), we can evaluate the local asymptotics of the Feynman Green
function by exploiting the Fock--Schwinger--DeWitt local asymptotics
for the heat kernel $K_{\mu \nu'}^{(1)}(\tau)$, i.e.
\begin{equation}
K_{\mu \nu'}^{(1)}(\tau) \sim {{\rm i}\over 16 \pi^{2}}
g^{{1\over 4}}\sqrt{\bigtriangleup} {g'}^{{1\over 4}}
{\rm e}^{{i\sigma \over 2\tau}}\sum_{n=0}^{\infty}
({\rm i}\tau)^{n-2}b_{n \; \mu \nu'},
\label{(6)}
\end{equation}
where $\sigma=\sigma(x,x')$, called the world function, is half the
square of the geodesic distance between $x$ and $x'$, the bi-scalar
$\sqrt{\bigtriangleup}(x,x')$ is defined by the equation
\begin{equation}
\sqrt{g} \; \bigtriangleup \; \sqrt{g'}
={\rm det} \; \sigma_{;\mu \nu'},
\label{(7)}
\end{equation}
and the coefficient bi-vectors $b_{n \; \mu \nu'}$ are evaluated,
in principle, from the recursion formula
\begin{equation}
\sigma^{; \lambda}b_{n \; \mu \nu';\lambda}
+n b_{n \; \mu \nu'}={1\over \sqrt{\bigtriangleup}}
\left(\sqrt{\bigtriangleup}b_{n-1,\mu \nu'}
\right)_{;\lambda}^{\; \; \; \lambda}
-R_{\mu}^{\; \; \lambda} \; b_{n-1,\lambda \nu'},
\label{(8)}
\end{equation}
obtained substituting the asymptotic expansion (6) in Eq. (3).
On using Eqs. (4)--(6) and the formulae (suitable restrictions
on the real parts of the parameters are taken to hold, to
ensure existence of our integrals~\cite{bi})
\begin{equation}
\int_{0}^{\infty}y^{-\beta}{\rm e}^{{\rm i}y}dy={\rm i}
\Gamma(1-\beta){\rm e}^{-{\rm i}{\pi \over 2}\beta},
\label{(9)}
\end{equation}
\begin{equation}
\int_{0}^{\infty}x^{\beta -1}\Gamma(\nu,cx)dx
={\Gamma(\beta+\nu)\over \beta c^{\beta}},
\label{(10)}
\end{equation}
we eventually find, in the limit as $\sigma(x,x') \rightarrow 0$,
the local asymptotics of the Feynman Green function in the form
\begin{equation}
G_{\mu \nu'}^{(\alpha)} \sim {{\rm i}\over 16 \pi^{2}}
\lim_{s \to 0}{\mu_{A}^{2s}\over \Gamma(s+1)}
{\cal G}_{\mu \nu'}^{(\alpha)}(s),
\label{(11)}
\end{equation}
where, after having defined
\begin{equation}
U_{n \; \mu}^{\; \; \; \; \lambda}(s;\alpha) \equiv
{2\over \sigma(x,x')}\delta_{\mu}^{\; \; \lambda}
+{(\alpha^{s+1}-1)\over (s+n)(s+1)}\nabla_{\mu}\nabla^{\lambda},
\label{(12)}
\end{equation}
\begin{equation}
B_{n \; \lambda \nu'}(s) \equiv b_{n \; \lambda \nu'}
\sqrt{\bigtriangleup}(x,x')(\sigma(x,x')/2)^{s+n},
\label{(13)}
\end{equation}
we write
\begin{equation}
{\cal G}_{\mu \nu'}^{(\alpha)}(s) \equiv \sum_{n=0}^{\infty}
\Gamma(1-s-n)U_{n \; \mu}^{\; \; \; \; \lambda}(s;\alpha)
B_{n \; \lambda \nu'}(s).
\label{(14)}
\end{equation}

\section{Singularities in the local asymptotics}

Since $b_{0 \; \mu \nu'}=g_{\mu \nu'}$, i.e. the parallel
displacement matrix, we recover immediately, from Eqs.
(11)--(14), the term
$$
{{\rm i}\over 8\pi^{2}}
{\sqrt{\bigtriangleup}(x,x')\over \sigma(x,x')
+{\rm i}\varepsilon}g_{\mu \nu'}
$$
in the local asymptotics of $G_{\mu \nu'}^{(\alpha)}$. Moreover,
on setting $\alpha=1$ for simplicity, the infinite sum (14) can be
evaluated with the help of the Euler--Maclaurin summation formula.
This provides, {\it among the others}, a term given by the integral
(the integer $n$ being replaced by the continuous variable $z$)
\begin{equation}
{J_{\mu \nu'}(s)\over \sqrt{\bigtriangleup}(x,x')} \sim
\log(\sigma(x,x')/2)\int_{0}^{y^{*}}
y \Gamma(-y-s)b_{y+1,\mu \nu'}dy,
\label{(15)}
\end{equation}
with $y^{*}$ in a small neighbourhood of the origin. On taking at
last the $s \rightarrow 0$ limit we therefore recover the familiar
$\log(\sigma(x,x'))$ singularity of the photon Green function, which
results, ultimately, from non-vanishing Riemann curvature.

\section{Hadamard function: second covariant derivatives in the
coincidence limit}

Our Hadamard function $G_{\mu \nu'}^{H}$ 
is the imaginary part of the Feynman Green
function in the formulae (11)--(14), and coincidence limits of
its second covariant derivatives can be used to find the regularized
energy-momentum tensor (see Christensen in Ref.~\cite{do}),
since (the subscript $+$ denoting anti-commutators)
\begin{equation}
G_{\mu \nu'}^{H}= \langle [A_{\mu},A_{\nu'}]_{+} \rangle ,
\label{(16)}
\end{equation}
\begin{eqnarray}
\langle F_{\rho \gamma} F_{\tau \beta} \rangle
&=&\lim_{x' \to x}{1\over 4}\Bigr(
G_{\gamma \beta'; \rho \tau'}^{H}
+G_{\beta \gamma'; \tau \rho'}^{H}
-G_{\gamma \tau'; \rho \beta'}^{H} \nonumber \\
&-& G_{\tau \gamma';\beta \rho'}^{H}
-G_{\rho \beta';\gamma \tau'}^{H}
-G_{\beta \rho'; \tau \gamma'}^{H} \nonumber \\
&+& G_{\rho \tau'; \gamma \beta'}^{H}
+G_{\tau \rho'; \beta \gamma'}^{H} \Bigr),
\label{(17)}
\end{eqnarray}
\begin{equation}
\langle T^{\mu \nu} \rangle_{\rm Maxwell}
=\lim_{x' \to x} \left[\biggr(g^{\mu \rho}g^{\nu \tau}
-{1\over 4}g^{\rho \tau}g^{\mu \nu}\biggr)
g^{\gamma \beta} \langle F_{\rho \gamma} F_{\tau \beta}
\rangle \right].
\label{(18)}
\end{equation}
In our investigation we find that the minimal-operator part of the 
Hadamard function contributes the divergent part (here square
brackets [...] denote the coincidence limits)
$$
\left(\Gamma(0) \left \{ \Bigr[b_{1 \; \gamma \beta';\rho \tau'}
\Bigr]-{1\over 6}\Bigr[b_{1 \; \gamma \beta'}\Bigr]R_{\rho \tau}
\right \} -{1\over 2}\Gamma(-1)\Bigr[b_{2 \; \gamma \beta'}\Bigr]
g_{\rho \tau}\right),
$$
while the non-minimal operator part of the Hadamard function
contributes further divergent terms given by
$$
(\alpha-1)\lim_{\varepsilon \to 0}\left({1\over \varepsilon}
\Gamma(1)A_{\beta \gamma \rho \tau}
+\Gamma(0)B_{\beta \gamma \rho \tau}
+{1\over 2}\Gamma(-1)C_{\beta \gamma \rho \tau}\right),
$$
where $\varepsilon$ is here just a device to deal with
${\delta_{n0}\over n}$ evaluated at $n=0$, and
\begin{eqnarray}
A_{\beta \gamma \rho \tau} & \equiv & \Bigr[
g_{\lambda \beta'; \; \; \gamma \rho \tau'}^{\; \; \; \; \; \; \lambda}
\Bigr]-{1\over 6}\Bigr[
g_{\lambda \beta'; \; \; \gamma}^{\; \; \; \; \; \; \lambda}\Bigr]
R_{\rho \tau} \nonumber \\
&+& {1\over 6}\left(\Bigr[
g_{\lambda \beta'; \; \; \tau'}^{\; \; \; \; \; \; \lambda}\Bigr]
R_{\gamma \rho}-\Bigr[
g_{\lambda \beta'; \; \; \rho}^{\; \; \; \; \; \; \lambda}\Bigr]
R_{\gamma \tau} 
+ \Bigr[g_{\lambda \beta';\gamma \tau'}
\Bigr]R_{\; \rho}^{\lambda}
-\Bigr[g_{\lambda \beta';\gamma \rho}\Bigr]R_{\; \; \tau}^{\lambda}
\right) \nonumber \\
&+& \left(\Bigr[g_{\lambda \beta';\rho \tau'}\Bigr]
{1\over 6}R_{\; \; \gamma}^{\lambda}
+\Bigr[\sqrt{\bigtriangleup}_{;\; \; \gamma \rho \tau'}^{\; \lambda}
\Bigr]g_{\lambda \beta}\right),
\label{(19)}
\end{eqnarray}
\begin{eqnarray}
B_{\beta \gamma \rho \tau}& \equiv & {1\over 2}
\left(\Bigr[
{\overline b}_{1 \; \lambda \beta'; \; \; \tau'}^{\; \; 
\; \; \; \; \; \; \; \lambda}
\Bigr]g_{\gamma \rho}-\Bigr[
{\overline b}_{1\; \lambda \beta'; \; \; \rho}^{\; \; \; 
\; \; \; \; \; \; \lambda}
\Bigr]g_{\gamma \tau}\right) \nonumber \\
&-& {1\over 2}g_{\rho \tau}\left(\Bigr[
{\overline b}_{1\; \lambda \beta'; \; \; \gamma}^{\; \; \; 
\; \; \; \; \; \; \lambda}
\Bigr]+{1\over 6}\Bigr[{\overline b}_{1 \; \lambda \beta'}\Bigr]
R_{\; \; \gamma}^{\lambda}\right) \nonumber \\
&-& {1\over 12}\left(\Bigr[{\overline b}_{1 \; \lambda \beta'}\Bigr]
(R_{\; \; \tau}^{\lambda}g_{\gamma \rho}
+R_{\; \; \rho}^{\lambda}g_{\gamma \tau})
+\Bigr[{\overline b}_{1 \; \tau \beta'}\Bigr]R_{\gamma \rho}
+\Bigr[{\overline b}_{1 \; \rho \beta'}\Bigr]
R_{\gamma \tau} \right) \nonumber \\
&+& {1\over 2}\left(\Bigr[
{\overline b}_{1\; \rho \beta';\gamma \tau'}\Bigr]
-\Bigr[{\overline b}_{1 \; \tau \beta';\gamma \rho}\Bigr]
\right) \nonumber \\
&-& {1\over 6}\left(\Bigr[
{\overline b}_{1\; \lambda \beta'}\Bigr]
(R_{\; \; \rho \gamma \tau}^{\lambda}
+R_{\; \; \tau \gamma \rho}^{\lambda})
+{1\over 2}\Bigr[{\overline b}_{1 \; \gamma \beta'}\Bigr]
R_{\rho \tau}\right),
\label{(20)}
\end{eqnarray}
\begin{equation}
C_{\beta \gamma \rho \tau} \equiv -2 \left(\Bigr[
{\overline b}_{2 \; \tau \beta'}\Bigr]g_{\gamma \rho}
+\Bigr[{\overline b}_{2 \; \rho \beta'}\Bigr]g_{\gamma \tau}
\right)-{1\over 2}\Bigr[{\overline b}_{2 \; \gamma \beta'}
\Bigr]g_{\rho \tau}.
\label{(21)}
\end{equation}
We have therefore provided a covariant isolation of divergences
resulting from every coincidence limit of second covariant
derivatives of the Hadamard Green function. In our derivation of
the results (19)--(21) we first sum over $n$, then take the
$s \rightarrow 0$ limit and eventually evaluate the coincidence
limit as $x' \rightarrow x$.

\section{Summary}

Our first original contribution is the expression (11)--(14) for
the asymptotic expansion of the Feynman photon Green function
at small values of the world function $\sigma(x,x')$. Our formulae
provide a novel way of expressing the familiar ${1\over \sigma}$
and $\log(\sigma)$ singularities in the local asymptotics; since
they do not rely on the introduction of a mass term in the action
functional, they are of particular interest for the reasons
described in Sec. 1.

Moreover, we have obtained a formula for the coincidence limit of
second covariant derivatives of the Hadamard function which can be
used, in principle and also in practice, to obtain the Maxwell
and gauge-fixing parts of the regularized energy-momentum tensor
for arbitrary values of the gauge parameter $\alpha$.
In manifolds without boundary, the work of Endo~\cite{en} has
shown that the trace anomaly resulting from the regularized
$T^{\mu \nu}$ has the coefficient of the $\nabla^{\mu}\nabla_{\mu}R$
term which depends on the gauge parameter $\alpha$. In manifolds
with boundary, the integration of such a total divergence does not
vanish, and further boundary invariants contribute to the regularized
energy-momentum tensor. Thus, the calculation expressed by (19)--(21)
is not of mere academic interest, but is going to prove especially
useful when boundary effects are included. Of course, physical 
predictions are expected to be independent of $\alpha$, but the
actual proof is then going to be hard. More precisely, the work
by Brown and Ottewill,~\cite{brot} which differs from our approach
because the $\alpha=1$ case is there considered and the 
${1\over \sigma}$ and $\log(\sigma)$ singularities in the propagator
are there assumed rather than derived, has been exploited by
Allen and Ottewill~\cite{alot} to show that, on using the Ward
identity and the ghost wave equation, the energy-momentum tensor
is $\alpha$-independent up to geometric terms (i.e. up to polynomial
expressions of dimension length$^{-4}$ formed from the metric, the
Riemann tensor and its covariant derivatives). The extension of these
results to manifolds with boundary is, to our knowledge, an open
research problem. 

\section*{Acknowledgments}
The work of G. Bimonte and G. Esposito has been partially 
supported by PRIN 2002 Sintesi. The INFN and Oklahoma University
financial support is gratefully acknowledged by G. Esposito.

\end{document}